\documentstyle[fleqn]{tp}

\input psfig

\newcommand\approxgt{\mbox{$^{>}\hspace{-0.24cm}_{\sim}$}}
\newcommand\approxlt{\mbox{$^{<}\hspace{-0.24cm}_{\sim}$}}

\begin{document}

\twocolumn[
\title{The Lyman-alpha Forest: a Cosmic Gold Mine}
\author{Lam Hui\\
{\it NASA/Fermilab Astrophysics Center,}\\
{\it Fermi National Accelerator Laboratory, Batavia, IL 60510}}
\vspace*{16pt}   

ABSTRACT.\
In recent years, a remarkably simple physical picture of the
Lyman-alpha forest has emerged, which allows detailed
predictions to be made and turns
the forest into a powerful probe of cosmology. 
We point out ways in which such a picture can be tested observationally, 
and explore three areas in which the Lyman-alpha forest can yield
valuable constraints: 
the reionization history, the primordial mass power spectrum and
the cosmological constant or its variants. The possibility of
combining with other high redshift observations, such
as the Lyman-break galaxy surveys, to provide
consistency checks and complementary information is also discussed.
\endabstract]

\markboth{Lam Hui}{The Lyman-alpha Forest: a Cosmic Gold Mine}

\small

\section{Introduction}
\label{intro}

The Ly$\alpha$ forest was predicted and observed in the 60s
(Gunn \& Peterson 1965, Bahcall \& Salpeter 1965, Lynds \& Stockton
1966, Burbidge et al. 1966, Kinman 1966; see Rauch 1998 for further
ref.). Since then, there have been many attempts to place
the study of the forest within the framework of cosmological structure
formation 
theories (e.g. Doroshkevich \& Shandarin 1980, Rees 1986, Bond et al.
1988, McGill 1990, Bi et al. 
1992). Recent numerical simulations lent support to some of these
ideas, while clarifying the nature of the forest and allowing
detailed predictions to be made (e.g. Cen et al. 1994,
Hernquist et al. 1995, Zhang et al. 1995, Petitjean et al. 1995,
Miralda-Escud\'e et al. 1996, Bond
\& Wadsley 1997, Theuns et al. 1998). 
We explain in this contribution the physical picture of the
forest that has emerged from these and other semi-analytical work (\S
\ref{basics}), discuss 
observational tests of this picture (\S \ref{conclude}), and point out
three examples of how such a picture allows us to ``mine'' valuable
cosmological information from the forest (\S \ref{mine}).

\section{Simplicity of the Forest}
\label{basics}

Numerical simulations indicate that the low column-density ($N_{\rm
HI} \, \approxlt \, 10^{14.5} {\rm 
cm^{-2}}$) Ly$\alpha$ forest (for redshift $z \, \approxgt \, 2$), which
occupies a large fraction of any  
given quasar absorption spectrum, arises from density (and
velocity) fluctuations of a smooth intergalactic medium. 
Regions of enhanced density (in redshift-space) in this smoothly
fluctuating medium naturally suffers
stronger absorption, creating features like
absorption lines, which cannot necessarily be interpreted as due to
discrete, well-isolated clouds, as in traditional
theories. 

The dynamics of such a medium is simple: gravitational instability
on large scales, and smoothing due to finite gas-pressure on small
scales (Reisenegger \& Miralda-Escud\'{e} 1995, Bi \& Davidsen 1997,
Hui, Gnedin \& Zhang 1997). Hence, the dark matter and baryon
components trace each other on large scales. The fluctuations giving rise to
the low $N_{\rm HI}$ 
forest are mildly nonlinear, with the overdensity $\delta
\rho/\bar\rho \, \approxlt \, 5$, where $\rho$ is the baryon density
and $\bar\rho$ its mean. The thermodynamics is well-understood.
Shock-heating is minimal and a tight temperature-density relation
exists as a result of three processes, recombination cooling,
photoionization heating and adiabatic heating/cooling: $T = T_0
\rho^{\gamma-1}$ where $T_0$ and $\gamma$ depend on reionization
history.

The optical depth $\tau$, which is related to the probability of
transmission ($f$) by $f = {\rm exp}[-\tau]$, is (e.g. Hui,
Gnedin \& Zhang 1997)
\begin{equation}
\tau (u_0) = \int A \rho^{\alpha} (x) W [u(x)-u_0] dx 
\label{tau}
\end{equation}
where $u_0$ is the velocity of observation, $x$ is the
comoving distance along the line of sight, $u$ is the total (Hubble
plus peculiar) velocity as a function of position, and
$W$ is the Voigt-profile given by $W[\Delta u] = (\pi b_T^2)^{-0.5}
{\rm exp}[-\Delta u^2/b_T^2]$ with $b_T = \sqrt{2 k_B T/m_p}$ where
$k_B$ is the Boltzman constant and $m_p$ is the proton mass.
The neutral hydrogen (which is what causes the Ly$\alpha$ absorption)
density is related to the baryon density $\rho$ by $A \rho^{\alpha}$,
where $1.6 \, \approxlt \, \alpha \, \approxlt \, 1.8$ for reasonable
reionization histories (Croft et al. 1997, Hui \& Gnedin 1997).
The constant $A$ depends on a number of parameters, among them the
cosmological mean baryon density, and the 
ionization background intensity $J$ whose size is uncertain by
an order of magnitude observationally. 

To predict the observational properties of the forest, all one needs
to know are then {\bf I.} the density and velocity distributions of the gas,
which can be obtained by hydrodynamic simulations (ref. in \S \ref{intro}),
analytical/semi-analytical approximations made possible by the mild
nonlinearity of 
the fluctuations (e.g. Reisenegger \& Miralda-Escud\'e 1995, Bi \&
Davidsen 1997, Gnedin \& Hui 1996, Hui et al. 1997)
or alternative efficient numerical techniques (e.g. Gnedin \& Hui 1998
\footnote{A regular Particle-Mesh N-body code can be modified
to compute the effective potential due to pressure in addition to
the usual gravitational potential, taking advantage of the known
temperature-density relation.}, Croft et
al. 1998); {\bf II.} the temperature-density relation which can
be obtained using semi-analytical methods (Hui \& Gnedin 1997); {\bf III.} the
constant $A$, which can be fixed by matching
the theoretical mean transmission with the observed one (Croft et al.
1998). 

Lastly, a few words on the 
higher column-density absorption systems: they
are likely to arise from more nonlinear objects such as collapsed
halos or galaxies. Such
objects seem to dominate the absorption spectra at low redshifts ($z
\, \approxlt \, 1$) (e.g. Lanzetta et al. 1995), but
not at higher redshifts (see Rauch 1998 for a review).

\section{The Cosmic Gold Mine}
\label{mine}
\subsection{Reionization History}
\label{reion}

Here we explore some effects of the reionization history
on a particular statistic called the b-distribution, which
has been measured by several groups (e.g. Hu et al. 1995, Lu et al.
1996, Kirkman \& Tytler 1997). 
The b-parameter refers to the width of a Voigt-profile-fit of an
absorption line. The number of lines as a function of width is
then the b-distribution. The picture outlined
in \S \ref{basics} provides a simple interpretation of the measured
widths. According to eq. (\ref{tau}), a peak of $\rho^{\alpha}$ in
velocity-space which is narrower than the thermal broadening width
$b_T$ will appear as a peak of $\tau$ with the shape of a Voigt-profile
and a width of $b_T$. Hence, the measured b-parameter provides a
direct indication of the temperature of the gas. On the other hand,
structure formation theories predict and allow
large-scale fluctuations where a peak of $\rho^{\alpha}$ is much wider
than the local $b_T$, which means according to eq. (\ref{tau}) the
corresponding peak in $\tau$ is no longer given by a Voigt-profile
shape and its width reflects more the scale of the fluctuation
rather than the temperature of the gas i.e. the measured
b can no longer be equated with the thermal broadening width. 

In more quantitative terms, we can always perform the
following expansion around a given absorption peak:
\begin{eqnarray}
\label{tauexpand}
&& \tau (u) = {\rm exp}\left[{{\rm ln}\,\tau(u)} \right] \\ \nonumber
&& \sim \tau(u_{\rm
p}) \, {\rm exp}\left[{{1\over 
2}[{\rm ln}\, \tau]'' (u-u_{\rm p})^2}\right] \, ,
\end{eqnarray}
where $u_{\rm p}$ is the velocity coordinate of the line center, and
the prime denotes differentiation with respect to $u$ and the second
derivative is evaluated at $u_{\rm p}$. The
first derivative vanishes because $\tau$ is at a local
extremum. This expansion gives none other
than the Voigt profile itself: $\propto {\rm exp} [-(u-u_{\rm p})^2/b^2]$. 
Hence, the fitted b-parameter would be $b = \sqrt{-2/[{\rm ln}
\tau]''}$.
Assuming most fitted-absorption lines do arise from peaks in $\tau$,
and assuming ${\rm ln} \tau$ is Gaussian random (as in the case
of linear fluctuations, or in the lognormal model), it can
be shown that the normalized distribution of lines is given by
(Hui \& Rutledge 1997)
\begin{equation}
dN/db = 4 (b_\sigma^4 /b^5) {\rm exp}[-b_\sigma^4/b^4]
\label{dNdb}
\end{equation}
where $b_\sigma^4 = 2/\langle ([{\rm ln} \tau]'')^2 \rangle$
and $\langle \rangle$ refers to ensemble-averaging.
This model naturally explains the salient features of the observed
b-distribution: a sharp lower cut-off and a long high-b tail,
which could be hard to explain in traditional theories of the forest
(see Fig. 18 of Bryan et al. 1998
for a comparison of the above $dN/db$ with observations). 
In other words, high-b-peaks
correspond to gentle 
fluctuations which are unsuppressed (hence the long tail), while
low-b-peaks are sharp fluctuations which  
are statistically rare.

This model also tells us what determines the b-distribution, through
the parameter $b_\sigma$: {\bf I.} the 
(dimensionless) average amplitude of the fluctuations (how nonlinear
the field is; see below),
and {\bf II.} the three
smoothing scales in the problem: the observation resolution, 
the average thermal broadening scale and the baryon-smoothing-scale
due to finite gas pressure; for high-quality Keck spectra, the first
is probably negligible, while the latter two are comparable.

How do the thermal broadening scale and the baryon-smoothing-scale
change with reionization history?
For a fixed redshift of observation (say $z = 3$), as one raises
the redshift of reionization, the temperature at $z=3$ becomes lower
(Hui \& Gnedin 1997)\footnote{$T$ is quite independent of $J_{\rm HI}$
as long as $H$ is sufficiently ionized, but does depend
somewhat on $J_{\rm HeII}$.}
and the thermal broadening scale becomes smaller.
The effect on the baryon-smoothing-scale is more subtle.
It is true that the Jeans scale, like the thermal broadening scale,
is proportional to $\sqrt T$, and so lowering the temperature lowers
both. However, as shown by Gnedin \& Hui (1998), the true
baryon-smoothing-scale is in fact not given by the Jeans scale, but
generally given by something smaller. In fact, right before
reionization occurs, the baryon-smoothing-scale is very small,
and afterwards, the baryon-smoothing-scale
does not suddenly jump up to the conventional Jeans scale, but only
slowly catches up with it.
This means that raising the redshift of reionization could have
the opposite effect of allowing a {\it larger} baryon-smoothing-scale
by allowing more time for it to catch up with the Jeans scale.
Therefore, exactly how raising the redshift of reionization affects
the b-distribution requires a detailed calculation.

This is a subject of much current interest, particularly
because of the work of Bryan et al. (1998) (see also Haehnelt \&
Steinmetz 1998, Theuns et al. 1998) who, after carefully 
investigating the effect of 
numerical resolution and box-size on the b-distribution, 
found that the canonical $\sigma_8 = 0.7$ SCDM
model predicts a b-distribution that has too many narrow lines
than is observed, if the universe reionizes by $z \sim 6$. 
The interesting questions are:
{\bf I.} changing the fluctuation amplitude will likely shift
the b-distribution, but which way will it go? - the 
arguments leading to eq. (\ref{dNdb}) indicate that
lowering the amplitude would move the b's up (because
lowering the amplitude means more suppression of sharp fluctuations
or narrow lines), but
as pointed out by Hui \& Rutledge (1998), nonlinear corrections
could reverse this trend;
{\bf II.} reionization history
will no doubt affect the b-distribution, but as pointed
out above, which direction it will go is not
obvious and requires a detailed calculation; {\bf III.} as noted by Hui \&
Gnedin (1997), the mean temperature of the intergalactic medium
increases with $\Omega_b$ (see also Bryan et al. 1998) and decreases
with $\Omega_m$; it would be interesting to see whether changing
these would fix the discrepancy of Bryan et al..

It is clear from the above discussion that the inference
on reionization history from the b-distribution will depend
on assumptions made about the cosmological density parameters
and the power spectrum normalization.
To isolate the effect of reionization
history from the effect of power spectrum normalization,
one possibility is to consider the lowest b's, which
according to our picture, should correspond to
the minimum size imposed by either the thermal-broadening
scale, or the baryon-smoothing-scale. However,
one should keep in mind the possible complication
that not all fitted absorption-lines arise from
peaks in $\tau$, e.g. some narrow lines might be introduced to
``fill in'' the wings of absorption peaks which do not necessarily have
Voigt-profile shapes, in which case the physical meaning of
the widths of these lines is unclear.
One should check for
this possibility using simulations, or even
develop other characterizations of the line-width
which are less prone to systematics of this sort.

\subsection{The Primordial Mass Power Spectrum}
\label{Pofk}

This is an area pioneered by Croft et al. (1998), who
showed that the linear mass power spectrum can be reliably
recovered from the power spectrum of the transmission.
The idea works as follows. The transmission power spectrum $P^f$
along the line of sight
and the three-dimensional mass power spectrum $P^\rho$ are related to
each other on large scales 
by (Hui 1998):
\begin{equation}
P^f (k_\parallel) = \int_{k_\parallel}^\infty B W(k_\parallel, k)
P^\rho (k) kdk 
\label{PfPrho}
\end{equation}
where $P^f$ is the fourier transform of the two-point correlation
$\langle \delta_f (0) \delta_f (u) \rangle$ ($\delta_f = (f - \bar f)/
\bar f$ where $\bar f$ is the mean transmission),
$k_\parallel$ is the velocity-wave-vector along the line of sight
and $k$ is the magnitude of the corresponding three-dimensional wave-vector.
The kernel $W(k_\parallel, k)$ describes the redshift-distortion
of the power spectrum. $B$ is a constant which
is determined by the nonlinear transformation from density to the
transmission ${\rm exp}(-\tau)$ (eq. [\ref{tau}]).
Fortunately, the only free-parameter that enters into the
determination of the constant $B$ is the constant $A$ in eq.
(\ref{tau}) which can be fixed by matching e.g. the observed mean
transmission (Croft et al. 1998). It is important to realize
that we have made use of the facts that $\alpha$ has a small range
(eq. [\ref{tau}]; \S \ref{basics}), that
the uncertain thermal-broadening and baryon-smoothing-scales
only affect the transmission correlation on small scales, 
and that the dark matter and baryon distributions trace each other on
large scales.

It is expected eq. (\ref{PfPrho}) holds on large
scales with
$P^\rho$ being the linear power spectrum. Hence, once the distortion
kernel is fixed (which is predicted by gravitational instability,
with dependence on the cosmological density parameters $\Omega$'s), 
the primordial mass power spectrum can be recovered from $P^f (k)$ by
inverting an essentially triangular matrix proportional to
the distortion kernel (Hui 1998). 

There are two main advantages of this way of measuring the
linear mass power spectrum. First, unlike in the case of galaxies,
the ``biasing'' relation between density and the observable (the
transmission) is known exactly here (aside from the parameter $A$
which can be fixed using independent observations). Second,
this provides a direct probe of the linear mass power spectrum on small
scales ($k \, \approxlt 0.02 {\rm s/km} \sim 4 h
\sqrt{\Omega_m} {\rm
Mpc^{-1}}$, for $z \sim 3$; see Croft et al. 1998), which are out of
the reach of
galaxy surveys at low redshifts (unless one corrects for the nonlinear
evolution). This is simply because the nonlinear scale becomes smaller
at higher redshifts.

Exciting (and beautiful!) first results of the application of the
above techniques 
to the observed forest were reported recently by Croft et al. (1998b),
Weinberg et al. (1998) (see also contribution to this volume by 
Weinberg). Here, let us 
list a few issues that deserve some thought and perhaps further investigation.
{\bf I.} The distortion kernel $W$ above can be predicted using
linear theory, but it is quite possible that the highly nonlinear
transformation from the density to the transmission will alter its
behavior, even at very large scales (McDonald \& Miralda-Escud\'{e}
1998, Hui 1998). This should
be carefully checked using simulations. 
{\bf II.} 
There is an upper
limit to the scale above which we 
cannot reliably recover the mass power spectrum. It is set by
the continuum, which is known to have long range fluctuations, and
which can only be estimated up to a limited accuracy. This limit
is around $k \sim 0.002 {\rm s/km}$, but should be checked carefully using
high resolution observations. {\bf III.} The whole inversion procedure
discussed above relies on the fact that eq. (\ref{tau}) holds for
most of any quasar spectrum, aside from regions of strong
absorption such as those caused by collapsed halos. It would be
useful to have an estimate of how such regions affect the
recovery procedure. {\bf IV.} Numerical simulations are typically used
to effectively fix the constant $B$ in eq. (\ref{PfPrho}). It will be
useful to have a study of the minimal box-size and resolution required
for the problem at hand. For instance, while one expects that
resolution can be sacrificed as long as one is interested in large
scale fluctuations, resolution does affect the fixing of the amplitude
$A$ using the observed mean transmission (Croft, priv.
comm.). Analytical methods to fix $B$ would be very useful.
{\bf V.} The temperature-density relation mentioned
in \S \ref{basics} is expected to have a scatter. It would be good to have an
idea by how much it could modify the transmission power spectrum,
or in other words, the ``biasing'' relation between density and
transmission.

\subsection{The Cosmological Energy Contents}
\label{lambda}

Here, we discuss a
version of a test proposed by Alcock \& Paczy\'nski (1979; AP
hereafter), which is particularly sensitive to the presence of the
cosmological constant $\Lambda$, or more generally, a component of
the cosmological energy contents which has negative pressure, let us
call it $Q$, with an equation of state $p = w \rho$ ($w < 0$).
The ideas presented here have been considered by several groups
recently (Croft 1998, Seljak 1998, McDonald \& Miralda-Escud\'{e} 1998, Hui,
Stebbins \& Burles 1998).
AP observed that an object placed at a cosmological distance would have a
definite relationship between its angular and redshift extents, which is
cosmology-dependent.  
Consider an object with mean redshift $z$, and angular size
$\theta$. Its transverse extent in velocity units is
\begin{equation}
u_\perp(\theta) = {H \over {1+z}} D_A (z) \theta \ .
\label{uperp}
\end{equation}
Here $H$ is the Hubble parameter at redshift $z$, and
$D_A (z)$ is the angular diameter distance (Weinberg 1972). 
For spherical objects the radial and transverse extents are equal, but more
generally if the object is squashed radially by a factor $\alpha_s$, the
radial extent is $u_\parallel \equiv{c\Delta z\over 1+z}=\alpha_s u_\perp$.
Here $c$ is the speed of light and $u_\perp,\,u_\parallel\ll c$ is assumed. 
This parameter $u_\perp/\theta$ depends on all the different cosmological
density parameters $\Omega_m$ (matter), $\Omega_k$ (curvature) and
$\Omega_Q$ (Q), but is particularly sensitive to $\Omega_Q$.
It is significantly lower for a 
$Q$-dominated universe (for $w\,\approxlt\,-1/3$) than for a
no-$Q$-universe.
(see Fig. 1 of Hui et al. 1998)
Qualitatively, one can understand this as follows.
In a $Q$-dominated universe, it is well
known that a significantly larger radial-comoving-volume is
associated with a given redshift range $\Delta z$, compared to a
no-$Q$-universe i.e. $c \Delta z / H(z)$ is larger for a $Q$-dominated
universe. Hence, the angular-extent of a spherical
object of a given redshift-extent $\Delta z$, would appear to 
be larger for a $Q$-universe: $\theta = c \Delta z / H(z) / D_A (z)$
would be larger i.e. a smaller $u_\perp(\theta)$.
It turns out for an open universe with no $Q$, $H(z)$ and $D_A(z)$ 
roughly balance each other to make $u_\perp(\theta)$ similar
to that of a flat-no-$Q$-universe.


In the case of the Ly$\alpha$ forest, the ``object'' 
to use is the two-point correlation function, whose
``shape'' ($\alpha_s$) is not
spherical because of redshift-anisotropy induced by 
peculiar motion. We cannot
observe the full three-dimensional correlation directly; instead
we can measure 
the one-dimensional correlation along a line of sight, and
the cross-correlation between two close-by lines of sight, or their Fourier
counterparts: the auto- and the cross-spectra.
A comparison of the two provides a new version of the AP test.
More precisely, given the observed transmission power spectrum,
one can invert eq. (\ref{PfPrho}) to obtain the mass power spectrum,
and then use the following equation to predict the
cross-spectrum $P^f_\times$ between two lines of sight with separation
$\theta$ ($P^f_\times$ is the fourier transform of the
cross-correlation between the two lines of sight $a$ and $b$:
$\langle \delta_f^a (0) \delta_f^b (u) \rangle$):
\begin{eqnarray}
\label{Pcross}
P^f_\times (k_\parallel, \theta) =&& \int_{k_\parallel}^\infty B
W(k_\parallel/k, k) \\ \nonumber && P^\rho (k) J_0 [k_\perp
u_\perp(\theta)]  kdk 
\end{eqnarray}
where $J_0$ is the spherical Bessel function.
This prediction is cosmology dependent mainly through the
parameter $u_\perp(\theta)$, and also through the distortion kernel 
$W(k_\parallel/k, k)$. The latter generally depends on an effective
bias parameter (Hui 1998), aside from the cosmological density
parameters of interests, which 
can fortunately be determined
from simulations because the exact ``biasing relation between density
and the observable (the transmission) is known (eq. [\ref{tau}]),
unlike in the case of galaxies (where the AP test has been
contemplated by other authors). A comparison of the
predicted with the observed cross-spectra, for a given observed
auto-spectrum, provides a version of the AP test: assuming the wrong
cosmology will result in a wrong prediction for the cross-spectrum.
Note that in this test, there is no need to fix the constant $B$
because it enters into the auto- and cross-spectra in the same way.

According to Hui et al. (1998), to reach a $4-\sigma$
level discrimination between e.g. the $\Omega_m = 0.3$ - $\Omega_k = 0.7$
universe and the $\Omega_m = 0.3$ - $\Omega_\Lambda =
0.7$ universe, only $25$ pairs of quasar spectra, at angular
separations $0.5' - 2'$, would be required.
There are roughly $10$ such pairs of quasars with existing
spectra at the above angular separations, or
slightly larger, and at $z \approxgt 1$ (see e.g. 
Crotts \& Fang 1998 \& ref. therein). Upcoming surveys such as 
the AAT 2dF and SDSS are expected to increase this number by at least
an order of magnitude, and to higher redshifts.

\section{Tests and More}
\label{conclude}

The cosmological utility of the (low column density) forest relies very much on
the simple relation between the optical depth or transmission
and the density+velocity fields, which arises from the fact that
{\bf I.} the distribution of the latter is determined
by gravitational instability alone on large scales, and by
baryon-smoothing on small scales (e.g. explosions do not
play an important dynamical role); {\bf II.} the ionizing
background $J$ does not have significant spatial fluctuations
on scales of interest (see \S \ref{Pofk}).
Here, we loosely refer to this whole set of assumptions as
the smooth-fluctuation-paradigm.
The ability of most current simulation-inspired work on the forest, which
makes use of the above assumptions, to explain the observed
b and column-density distributions should be
counted as a vindication of the paradigm. After all, this body of work
is based on
structure formation models constructed to
match observations other than those of the forest.
Moreover, there are theoretical reasons to believe that $J$
fluctuations should be small on the scales we are interested in 
(e.g. Croft et al. 1998b).
Nonetheless, it is important that we find alternative
ways to test our assumptions observationally, as we continue
to look for new applications of the forest in cosmology.

Here, we discuss three possible tests that involve only the forest,
and one cross-test with another high redshift observation.

{\bf I.} Use double (or multiple) lines of sight. There are in fact
two different tests here. First, 
for quasar pairs that
are very close together e.g. lensed 
pairs, one can check whether the cross-correlation between the forest in
the two lines of sight is as strong as one would expect based
on the smooth-fluctuation-paradigm. More specifically, for lensed quasar
pairs which are typically of the order of arc-second separation,
the distance between the two lines of sight is sub-kpc,
which is much smaller than the baryon-smoothing-scale expected
for a gas of $T \sim 10^4 K$. This means one expects $100 \%$
correlation between these two lines of sight through the forest. Exactly
such a behavior 
has been observed by Rauch (1997). This shows there is no fluctuation in
the ionizing background, nor is there complicated motion in the forest due to
explosions, at least on 
these small scales. For quasar-spectra at larger separations, on the other
hand, assuming a given set of $\Omega$'s, one can turn the procedure
in \S \ref{lambda} around, and test for the redshift-anisotropy due to
peculiar motion. This is a robust prediction of gravitational
instability. Any other source of fluctuations, such as that due
to a fluctuating background, will act as additional sources of
``biasing'', and change the prediction for the redshift-anisotropy.

{\bf II.} Use higher order statistics. It is well known that gravitational
instability gives robust predictions for higher order statistics such
as the skewness $\langle \delta^3 \rangle / \langle \delta^2 \rangle$
(Peebles 1980), where $\delta$ is the overdensity in mass.
Just as in the case of galaxies, where measuring the skewness for the
galaxy distribution ($\langle \delta_g^3 \rangle / \langle \delta_g^2
\rangle$ where $\delta_g$ is the galaxy overdensity) provides
information on the biasing relation between mass and
galaxy distributions, measuring the skewness of the transmission should
provide a test of the ``biasing'' relation between transmission and
mass density (eq. [\ref{tau}]). Any additional complication due
to say a fluctuating ionizing background is going to change that
relation and hence the prediction for the higher moments.
Efforts are under way to calculate these quantities.
As far as measurements are concerned, it is important to
guard against possible estimation-biases (Hui \& Gazta\~{n}aga 1998).
It might also well be the case that other higher order measures
are better suited for the purpose at hand, because of the
particular highly nonlinear transformation from density to
transmission we have here (especially the exponentiation: ${\rm exp}(-\tau)$).

{\bf III.} If supernova explosions play an important role in the
distribution of the gas that makes up the forest, we should be
able to see their remnants: metals. Clever efforts to detect
metals in $N_{\rm HI} < 10^{14} {\rm cm^{-2}}$ forest have so far
only yielded upper limits (see Lu et al. 1998). However,
a carbon-to-hydrogen ratio of the order of $10^{-2.5}$ has been
detected in systems with $N_{\rm HI} \sim 10^{14.5} {\rm cm^{-2}}$.
It is unclear, though, what the corresponding volume-filling
factor of these slightly-metal-enriched regions is.
This aspect of the forest certainly deserves closer scrutiny.
\footnote{Note added after conference: Cowie \& Songaila (1998)
recently reported results that are in apparent contradiction
with Lu et al. (1998). If the results hold up, this implies
some amount of star formation occured at early times, but has stopped
by $z \sim 3$ in order not to violate point {\bf I} above.
Its impact on the forest remains to be investigated.}

Finally, it is appropriate at this conference to make some connection
with another probe of the high-redshift universe: the
Lyman-break-galaxy survey (Steidel et al. 1998). As noted by several
authors 
(Nusser \& Haehnelt 1998, McDonald \& Miralda-Escud\'e 1998, Croft et
al. 1998), information obtained from the Lyman-alpha forest
can be fruitfully combined with observations of the Lyman-break
objects, which are observed at similar redshifts ($z \sim 3$).
For instance, once the mass power spectrum is inferred
from the forest, a comparison with the clustering of
the Lyman-break objects would immediately yield a measure
of the bias of the Lyman-break galaxies. 
On the other hand, the bias of the Lyman-break galaxies can be obtained
from observations of the Lyman-break objects themselves: either by
a direct measurement of their masses (using high resolution infrared
observations, which should be feasible in the near future), or by
measuring the skewness of the galaxy-distribution. 
Note that to deduce from either quantity the bias (as defined by the
square root of the ratio
of the galaxy power spectrum to mass power spectrum), one needs
a model for how these galaxies form i.e. how they are formed
in relation to dark matter halos (Mo et al. 1998).

With the above information in hand, we have two independent 
measures of the mass power spectrum at $z \sim 3$, one from
the Ly$\alpha$ forest, the other from the Lyman-break-objects making
use of the deduced bias parameter. They probe the clustering
at different scales, and so one can use them to constrain
the shape of the power spectrum; or, if one assumes a particular
cosmological model with a definite power spectrum shape,
one can test for consistency. Moreover, the latter approach
also allows two independent measurements of $\Omega_m$ 
(with minor dependence on the other
parameters), by comparison with the cluster-normalized mass
power spectrum today (see e.g. Giavalisco et al. 1998,
Adelberger et al. 1998, Weinberg 1998).
Investigations along these lines are being pursued.

\section*{Acknowledgments}
I am grateful to Scott Burles, Nick Gnedin, Robert Rutledge, Albert Stebbins
and Yu Zhang for collaboration, some results of which are
reported here. I thank Mike Norman for useful discussions on 
the b-distribution, and thank Simon White and the Max-Planck Institut
fuer Astrophysik for hospitality. Support by the DOE and the NASA grant
NAG 5-7092 at Fermilab is gratefully acknowledged.




\end{document}